\documentstyle{article}

\input amssym.def
\input amssym

\newcommand{\ms}{\medskip}

\newcommand{\noi}{\noindent}
\newcommand{\ra}{\rightarrow}
\newcommand{\bea}{\begin{eqnarray}}
\newcommand{\eea}{\end{eqnarray}}

\newcommand{\gr}{Groenewold}
\newcommand{\vh}{Van~Hove}
\newcommand{\vn}{Von~Neumann}
\newcommand{\q}{{\cal Q}}
\newcommand{\p}{{\cal P}}

\newcommand{\f}{{\cal F}}

\newcommand{\bb}{{\cal B}}
\newcommand{\oo}{{\cal O}}

\newcommand{\nn}{{\cal N}}
\newcommand{\T}{{\theta}}
\newcommand{\va}{V_{\alpha}}
\newcommand{\wa}{W_{\alpha}}

\def\r{{\Bbb R}}
\def\z{{\Bbb Z}}
\def\c{{\Bbb C}}

\newsymbol\circledS 1073
\newsymbol\ltimes 226E

\newtheorem{thm}{Theorem}
\newtheorem{lem}{Lemma}
\newtheorem{cor}[thm]{Corollary}
\newtheorem{prop}[thm]{Proposition}
\newtheorem{defn}{Definition}

\def\f #1,#2.{\textstyle{#1\over #2}}

\def\sp{{\rm span}}

\title{On Quantizing $T^*\!S^1$}

\author{{\bf Mark J. Gotay}\thanks{Supported in part by NSF grant
DMS 96-23083.}
\\ Department of Mathematics \\ University of Hawai`i \\ 2565 The
Mall \\ Honolulu, HI 96822 USA \\
\and {\bf Hendrik B. Grundling} \\ Department of Pure Mathematics
\\ University of New South Wales \\ P.O. Box 1 \\ Kensington, NSW
2033 Australia}

\date{September 23, 1996}

\begin{document}

%%%%%%%%%%%%%%%%%%%%%%%%%%%%%%%%%%%%%%%%%%%%%%%%%%%%%%%%%%%%%%%%%%%%%%
%%%%%%%%%%%%%%%%%%%%%%%%%%%%%%%%%%%%%%%%%%%%%%%%%%%%%%%%%%%%%%%%%%%%%%
%%%%%%%%%%%%%%%%%%%%%%%%%%%%%%%%%%%%%%%%%%%%%%%%%%%%%%%%%%%%%%%%%%%%%%
%%%%%%%%%%%%%%%%%%%%%%%%%%%%%%%%%%%%%%%%%%%%%%%%%%%%%%%%%%%%%%%%%%%%%%
%%%%%%%%%%%%%%%%%%%%%%%%%%%%%%%%%%%%%%%%%%%%%%%%%%%%%%%%%%%%%%%%%%%%%%
%%%%%%%%%%%%%%%%%%%%%%%%%%%%%%%%%%%%%%%%%%%%%%%%%%%%%%%%%%%%%%%%%%%%%% 

\maketitle

\begin{abstract} In this paper we continue our study of \gr-\vh\
obstructions to quantization. We show that there exists such an
obstruction to quantizing the cylinder $T^*\!S^1.$ More precisely,
we prove that there is no quantization of the Poisson algebra of
$T^*\!S^1$ which is irreducible on a naturally defined
$\mbox{e(2)} \times \r$ subalgebra. Furthermore, we
determine the maximal ``polynomial'' subalgebras that can be
consistently quantized, and completely characterize the
quantizations thereof. This example provides support for one of the
conjectures in
\cite{GGT}, but disproves part of another. Passing to coverings, we
also derive a no-go result for $\r^2$ which is comparatively
stronger than those originally found by \gr\ \cite{Gr} and \vh\
\cite{vH1}. \end{abstract}

%%%%%%%%%%%%%%%%%%%%%%%%%%%%%%%%%%%%%%%%%%%%%%%%%%%%%%%%%%%%%%%%%%%%%%
%%%%%%%%%%%%%%%%%%%%%%%%%%%%%%%%%%%%%%%%%%%%%%%%%%%%%%%%%%%%%%%%%%%%%%
%%%%%%%%%%%%%%%%%%%%%%%%%%%%%%%%%%%%%%%%%%%%%%%%%%%%%%%%%%%%%%%%%%%%%%
%%%%%%%%%%%%%%%%%%%%%%%%%%%%%%%%%%%%%%%%%%%%%%%%%%%%%%%%%%%%%%%%%%%%%%
%%%%%%%%%%%%%%%%%%%%%%%%%%%%%%%%%%%%%%%%%%%%%%%%%%%%%%%%%%%%%%%%%%%%%%
%%%%%%%%%%%%%%%%%%%%%%%%%%%%%%%%%%%%%%%%%%%%%%%%%%%%%%%%%%%%%%%%%%%%%% 

\begin{section}{Introduction}

Let $M$ be a symplectic manifold and $\p(M)$ its associated Poisson
algebra. In \cite{GGT} we conjectured that: 

\begin{quote}
\em Let\/
$\bb \subset \p(M)$ be a ``basic set'' of observables, such that
the Poisson algebra it generates is finite-dimensional. Then there
is no nontrivial strong quantization of\/
$\big(\p(M),\bb\big)$.
\end{quote}

\noi To understand this, we recall the basic definitions;
motivation for these can be found in \cite{GGT}.

\begin{defn}$\,\,${\rm A {\sl basic set} of observables $\bb$ is a
linear subspace of $\p(M)$ such that: \begin{itemize}
\begin{itemize}
\item $\bb$ is finite-dimensional, \vskip 6pt
\item the Hamiltonian vector fields $X_f$, $f\in\bb$, are
complete, \vskip 6pt
\item $\{X_f\,|\,f\in\bb\}$ span the tangent spaces to $M$
everywhere, \vskip 6pt
\item $1\in\bb$, and
\vskip 6pt
\item $\bb$ is a minimal space satisfying these requirements.
\vskip 6pt
\end{itemize}
\end{itemize}	}
\end{defn}

A basic set typically consists of the components of the momentum
map for a transitive Hamiltonian action of a finite-dimensional Lie
group on $M$. 

\begin{defn}$\,\,${\rm Let $\bb$ be a basic set, and let $\oo$ be a
Poisson subalgebra of $\p(M)$ containing $\bb$. Then a {\sl
quantization} of $(\oo,\,\bb)$ is a linear map $\q$ from $\oo$ to
the algebra of symmetric operators which preserve some fixed domain
$D$ in some Hilbert space, such that for all $f,\,g\in\oo$,
\begin{itemize}
\begin{itemize}
\item $\q(\{f,\,g\})={i\over\hbar}\big[\q(f),\,\q(g)\big]$,
\vskip 6pt
\item $\q(1)=I$,
\vskip 6pt
\item if $X_f$ is complete, then $\q(f)$ is essentially
self-adjoint on $D$,
\vskip 6pt
\item $\q(\bb)=\big\{\q(f)\,|\,f\in\bb\big\}$ is an
irreducible set, and \vskip 6pt
\item $D$ contains a dense set of separately analytic vectors
for some basis of $\q(\bb)$.
\vskip 6pt
\end{itemize}
\end{itemize} 
A quantization $\q$ is {\sl strong\/} if in addition $D$ contains a
dense set of separately analytic vectors
for some Lie generating basis of $\q\big({\cal N}_{\oo}(\wp
(\bb))\big)$, where ${\cal N}_{\oo}(\wp
(\bb))$ denotes the normalizer in $\oo$ of the Poisson subalgebra
$\wp(\bb)$ generated by $\bb$. Finally, a quantization is {\sl
trivial\/} whenever its representation space is zero- or
one-dimensional.}
\end{defn}

All examples which have been analyzed to date validate the
conjecture above; in particular, $\r^{2n}$ with the basic set \[\bb
= \sp\{1,q^i,p_i\,|\,i = 1,\ldots,n\}\] 

\noi \cite{Gr,vH1}, and $S^2$ with
\[\bb = \sp\{1,S_1,S_2,S_3\},\]

\noi where the $S_i$ are the components of the spin vector
\cite{GGH}. In both cases the basic sets are already Poisson
subalgebras. On the other hand, there does exist a nontrivial
strong quantization of $T^2$ with any of the basic sets
\[\bb_k = \sp\{1, \sin k\T, \cos k\T,\sin k\phi, \cos k\phi\}\] 

\noi for $k$ a positive integer \cite{Go}. But the Poisson algebras
generated by the $\bb_k$ are infinite-dimensional. In a sense,
$\bb_1$ is the toral analogue of the basic set for $\r^{2}$. 

Given this dichotomy, a natural example with which to test the
conjecture is the cylinder $T^*\!S^1$, since it is topologically
``halfway'' between $\r^2$ and $T^2$. Endow the cylinder with the
canonical Poisson bracket
\[\{f,g\} = \frac{\partial f}{\partial \ell}\frac{\partial
g}{\partial \T} - \frac{\partial f}{\partial \T}\frac{\partial
g}{\partial \ell},\] 

\noi where $\ell$ is the angular momentum conjugate to $\T$.
Although the symplectic self-action of
$T^*\!S^1$ is not Hamiltonian (thinking of
$T^*\!S^1$ as $T \times
\r$, where $T$ is the circle group), the cylinder can nonetheless be
realized as a coadjoint orbit of the Euclidean group E(2). The
corresponding momentum map
$T^*\!S^1 \ra \mbox{e}(2)^*$ has components $\{\ell, \sin \T, \cos
\T
\}$. Together with the constant function 1, these components span
the basic set

\[\bb = \sp\{1, \ell, \sin \T, \cos \T\}.\] 

\noi Algebraically, $\bb$ is the cylindrical analogue of those for
$\r^{2}$ and $T^2$.

In this paper we show that the conjecture holds for the cylinder:
there is no quantization of $\p(T^*\!S^1)$--strong or
otherwise--which is irreducible when restricted to $\bb$. However,
there do exist quantizations of certain ``polynomial'' subalgebras
of
$\p(T^*\!S^1)$; these are discussed and characterized in \S3.
Finally, we ``lift'' our results to $\r^2$, thereby producing a
no-go result which is the strongest yet obtained for $\r^2.$

\end{section}

%%%%%%%%%%%%%%%%%%%%%%%%%%%%%%%%%%%%%%%%%%%%%%%%%%%%%%%%%%%%%%%%%%%%%%
%%%%%%%%%%%%%%%%%%%%%%%%%%%%%%%%%%%%%%%%%%%%%%%%%%%%%%%%%%%%%%%%%%%%%%
%%%%%%%%%%%%%%%%%%%%%%%%%%%%%%%%%%%%%%%%%%%%%%%%%%%%%%%%%%%%%%%%%%%%%%
%%%%%%%%%%%%%%%%%%%%%%%%%%%%%%%%%%%%%%%%%%%%%%%%%%%%%%%%%%%%%%%%%%%%%%
%%%%%%%%%%%%%%%%%%%%%%%%%%%%%%%%%%%%%%%%%%%%%%%%%%%%%%%%%%%%%%%%%%%%%%
%%%%%%%%%%%%%%%%%%%%%%%%%%%%%%%%%%%%%%%%%%%%%%%%%%%%%%%%%%%%%%%%%%%%%% 

\begin{section}{The Obstruction}

Our first task is to determine all possible quantizations of the
basic set
$\bb \cong \mbox{e(2)} \times \r$. According to the definition a
quantization of
$\bb$ in this instance amounts to a Lie algebra representation $\q$
by essentially self-adjoint operators on a common invariant dense
domain in a Hilbert space which is both irreducible and integrable.
Thus it suffices to compute the derived representations
corresponding to the unitary irreducible representations (``UIRs'')
of the universal covering group of
${\mbox{E(2)}} \times \r$.

Now, the universal covering group of E(2) is the semi-direct
product $\r \ltimes \r^2$ with the composition law \[(t,x,y)\cdot
(t',x',y') = (t+t', x' \cos t + y' \sin t + x, y' \cos t - x' \sin
t + y).\]

\noi Fortunately, it is straightforward to determine the UIRs of
this group. {}From the theory of induced representations of
semi-direct products \cite{Ma} (see also \cite[\S5.8]{Is}), we
compute that these representations are of two types:
\begin{enumerate}
\begin{enumerate}
\item[({\em i\/})] $\big(U(t,x,y)\psi\big)(\T) = e^{i\lambda(x \cos
\T + y
\sin \T)}e^{i\nu t}\psi(\T + t)$ on
$L^2(S^1)$, and
\vskip 8pt
\item[({\em ii\/})] $U(t,x,y)z = e^{i\mu t}z$ on $\c$. \vskip 8pt
\end{enumerate}
\end{enumerate}

\noi Here $\lambda, \nu, \mu$ are real parameters satisfying
$\lambda > 0$ and $0 \leq \nu < 1.$ The corresponding derived
representations are\footnote{We denote multiplication operators as
functions.} 

\begin{enumerate}
\begin{enumerate}
\item[({\em i\/})$'$] $\q(\ell) = -i\,{\displaystyle \frac{d}{d\T}}
+ \nu I$,
$\q(\sin
\T) = \lambda \sin \T$, $\q(\cos \T) = \lambda \cos \T$
\vspace{2ex} on \linebreak[4]
$C^{\infty}(S^1,\c) \subset L^2(S^1)$, and \vskip 8pt
\item[({\em ii\/})$'$] $\q(\ell) = \mu$, $\q(\sin \T) = 0$,
$\q(\cos \T) = 0$ on $\c$.
\vskip 8pt
\end{enumerate}
\end{enumerate}
\noi Thus the required representations of ${\mbox{e(2)}}
\times \r$ are given by ({\em i\/})$'$ and ({\em ii\/})$'$
supplemented by the condition $\q(1) = I.$ The parameter $\lambda$
can be identified with the reciprocal of Planck's reduced
constant,\footnote{See
\cite[\S4.6]{Is} for a discussion. There is an error here,
however; $\lambda$ should be identified with $\hbar^{-1}$, and not
$\hbar^2$.} which we take to be one.

\ms

Rather than consider the entire Poisson algebra $\p(T^*\!S^1)$,
we will focus on the Poisson subalgebra $P$ of polynomials in
elements of $\bb$, i.e., sum of multiples of terms of the form
\[\ell^{\, r} \sin^m\!\T \cos^n\!\T\]

\noi with $r,m,n$ nonnegative integers. We remark in passing that
$P$ is dense in $C^{\infty}(T^*\!S^1)$, where the latter is given
the topology of uniform convergence on compacta of a function and
its derivatives. Let
$P^r$ be the subspace thereof consisting of polynomials which are
at most degree $r$ in $\ell,$ and
$P_r$ those which are homogeneous of degree $r$ in $\ell.$ A short
calculation shows that the normalizer of $\bb$ in both $P$ and
$\p(T^*\!S^1)$ is just itself, so that any quantization of
$(P,\bb)$ or $\big(\p(T^*\!S^1),\bb\big)$ is automatically strong.

\ms

Now suppose there existed a quantization $\q$ of $(P,\bb)$ on some
common invariant dense domain $D$ in an infinite-dimensional
Hilbert space $\cal H$. Arguing as in the proof of
\cite[Proposition 2]{GGT}, we may assume that $D =
C^{\infty}(S^1,\c)$ in ${\cal H} = L^2(S^1)$, so that $\q$
restricted to
$\bb$ is given by ({\em i\/})$'$. To begin, we generate some ``\vn\
rules.''

\begin{prop}
$\q(\ell^{\, 2}) = \q(\ell)^2 + b\q(\ell) + cI$, where $b,c \in \r$
are arbitrary.
\label{prop:l2}
\end{prop}

\noi{\em Proof:} First observe that $L^2(S^1)$ has an orthonormal
basis $\{|n\rangle \,|\,n \in \z\}$ of eigenvectors of $\q(\ell)$,
where $|n\rangle = \frac{1}{\sqrt{2\pi}}e^{in\T}$. Thus $\q(\ell)$
has spectrum $\{n + \nu\,|\, n \in \z\}$, and the multiplicity of
each eigenvalue is 1. Hence, if a normal operator commutes with
$\q(\ell)$ on
$C^{\infty}(S^1,\c)$, then it must be a function of $\q(\ell)$.

Also observe that
\begin{equation}
\q(\cos \T)|n\rangle = \frac{1}{2}\big(|n+1\rangle +
|n-1\rangle\big) \label{eq:c}
\end{equation}
\noi and
\begin{equation}
\q(\sin \T)|n\rangle = \frac{1}{2i}\big(|n+1\rangle -
|n-1\rangle\big). \label{eq:s}
\end{equation}

Now set $\Delta = \q(\ell^{\, 2}) - \q(\ell)^2$. Then
$[\Delta,\q(\ell)] = 0$, so that $\Delta$ is a function of
$\q(\ell)$, say $\Delta = \xi\big(\q(\ell)\big)$, and we want to
compute $\xi$. Since \begin{equation}
\xi\big(\q(\ell)\big) |n\rangle = \xi(n+\nu) |n\rangle \label{eq:xi}
\end{equation}

\noi and $\{|n\rangle\}$ span
$L^2(S^1)$, it suffices to determine the sequence
$\{\xi(n+\nu)\,|\,n \in \z\}$.

Let us quantize the Poisson bracket identity \begin{equation}
\big\{\{\ell^{\, 2},\sin \T\},\sin \T\big\} + \big\{\{\ell ^{\,
2},\cos
\T\},\cos \T\big\} = 2
\label{eq:iden}
\end{equation}

\noi to get
\[\big[[\q(\ell^{\, 2}),\q(\sin \T)],\q(\sin \T)\big] +
\big[[\q(\ell^{\, 2}),\q(\cos \T)],\q(\cos \T)\big] = - 2,\]

\noi whilst explicit calculation produces
\[\big[[\q(\ell)^2,\q(\sin \T)],\q(\sin \T)\big] +
\big[[\q(\ell)^2,\q(\cos \T)],\q(\cos \T)\big] = - 2.\] 

\noi Subtracting,
\[\big[[\Delta,\q(\sin \T)],\q(\sin \T)\big] + \big[[\Delta,\q(\cos
\T)],\q(\cos \T)\big] = 0.\] 

\noi Denote the left hand side of this equation by $K$. Now
evaluate the matrix element
$\langle n|K|n\rangle$ by substituting $\Delta =
\xi\big(\q(\ell)\big)$. After a short computation using
(\ref{eq:c})--(\ref{eq:xi}), we obtain the recursion relation
\[2\xi(n') - \xi(n'+1) - \xi(n'-1) = 0,\] 

\noi where $n' = n+ \nu.$ This has the solutions $\xi(n') = bn' +
c,$ where
$b,c$ are real as $\Delta$ is symmetric. Thus \[\Delta =
\xi\big(\q(\ell)\big) = b\q(\ell) + cI,\] 

\noi which yields the desired result. $\Box$ 

\vskip 12pt

Next, we quantize the relations
\[\ell \sin \T = -\frac{1}{2}\{\ell^{\, 2},\cos
\T\}\;\;\;\;\mbox{and}\;\;
\;\;\ell \cos \T = \frac{1}{2}\{\ell^{\, 2},\sin \T\}\] 

\noi thereby obtaining
\begin{equation}
\q(\ell \sin \T) = \q(\sin \T)\q(\ell) - \frac{i}{2}\q(\cos \T)
+ \frac{b}{2}\q(\sin \T) \label{eq:ls}
\end{equation}

\noi and
\begin{equation}
\q(\ell \cos \T) =
\q(\cos \T)\q(\ell) + \frac{i}{2}\q(\sin \T) + \frac{b}{2}\q(\cos
\T). \label{eq:lc}
\end{equation}

\noi Then, quantizing the relation $\{\ell \cos \T,\ell \sin \T\} =
\ell$, we conclude that $b = 0$ in the above. Finally, using
(\ref{eq:ls}), (\ref{eq:lc}) and Proposition \ref{prop:l2}, we
quantize \[\ell^{\, 2} \sin \T = \frac{1}{2}\{\ell \cos \T,\ell^{\,
2}\}\;\;\;\; \mbox{and}\;\;\;\;\ell^{\, 2} \cos \T =
-\frac{1}{2}\{\ell \sin \T,\ell^{\, 2}\}\]

\noi to get
\begin{equation}
\q(\ell^{\, 2} \sin \T) =
\q(\sin \T)\q(\ell)^2 -i\q(\cos \T)\q(\ell) +
\frac{1}{4}\q(\sin \T)
\label{eq:l2s}
\end{equation}

\noi and
\begin{equation}
\q(\ell^{\, 2} \cos \T) =
\q(\cos \T)\q(\ell)^2 + i\q(\sin \T)\q(\ell) + \frac{1}{4}\q(\cos
\T).
\label{eq:l2c}
\end{equation}

Our main result is the following no-go theorem: 

\begin{thm} There is no nontrivial quantization of $(P,\bb)$.
\end{thm}

\noi {\em Proof:} We merely use ({\em i\/})$'$ and the \vn\ rules
(\ref{eq:l2s}) and (\ref{eq:l2c}) to quantize the
bracket relation
\[2\left\{\{\ell^{\, 2} \sin \T,\ell^{\, 2} \cos \T\}, \cos
\T\right\} = 12\ell^{\, 2} \sin \T.\]

\noi After simplifying, the left hand side reduces to
\[ 12\q(\sin \T)\q(\ell)^2 - 12i\q(\cos \T)\q(\ell) + 5\q(\sin
\T),\]

\noi whereas the right hand side is
\[ 12\q(\sin \T)\q(\ell)^2 - 12i\q(\cos \T)\q(\ell) + 3\q(\sin
\T),\]

\noi and the required contradiction is evident. $\Box$

\vskip 12pt

This theorem holds for representations of type ({\em i\/})$'$. But
it is easy to see that there are no trivial quantizations of
$(P,\bb)$ either, corresponding to representations of type ({\em
ii\/})$'$. Indeed, quantizing
$\cos^2 \T = \frac{1}{2}\big\{\{\ell^{\, 2},\sin \T\},\sin
\T\big\}$ we obtain $\q(\cos^2 \T) = 0.$ Likewise, $\q(\sin^2 \T) =
0.$ But this is impossible:\[I = \q(1) = \q(\cos^2 \T + \sin^2 \T)
= \q(\cos^2 \T) + \q(\sin^2 \T) = 0.\]

Assembling the above results, we therefore have 

\begin{cor} There exists no quantization of
$\big(\p(T^*\!S^1),\bb\big)$.
\label{cor:dne}
\end{cor}

Thus $\big(\p(T^*\!S^1),\bb\big)$ does indeed satisfy the
conjecture of \S1.

\end{section}

%%%%%%%%%%%%%%%%%%%%%%%%%%%%%%%%%%%%%%%%%%%%%%%%%%%%%%%%%%%%%%%%%%%%%%
%%%%%%%%%%%%%%%%%%%%%%%%%%%%%%%%%%%%%%%%%%%%%%%%%%%%%%%%%%%%%%%%%%%%%%
%%%%%%%%%%%%%%%%%%%%%%%%%%%%%%%%%%%%%%%%%%%%%%%%%%%%%%%%%%%%%%%%%%%%%%
%%%%%%%%%%%%%%%%%%%%%%%%%%%%%%%%%%%%%%%%%%%%%%%%%%%%%%%%%%%%%%%%%%%%%%
%%%%%%%%%%%%%%%%%%%%%%%%%%%%%%%%%%%%%%%%%%%%%%%%%%%%%%%%%%%%%%%%%%%%%%
%%%%%%%%%%%%%%%%%%%%%%%%%%%%%%%%%%%%%%%%%%%%%%%%%%%%%%%%%%%%%%%%%%%%%% 

\begin{section}{Quantizable Subalgebras of Observables} 

In view of the impossibility of quantizing $(P,\bb)$, one can ask
for the maximal subalgebras in $P$ to which we can extend an
irreducible representation of $\bb$.

Such subalgebras certainly exist: For instance, there is a two
parameter family of quantizations of the pair
$(P^1,\bb)$. They are the ``position representations'' on
$C^{\infty}(S^1,\c) \subset L^2(S^1)$ given by 
\begin{equation}
\q_{\nu,\eta}\big(f(\T)\ell + g(\T)\big) = -if(\T)\frac{d}{d\T} +
\left[\left(\eta - \frac{i}{2}\right)f'(\T) + \nu f(\T) + g(\T)
\right],
\label{eq:qnu}
\end{equation}

\noi where $\nu$ labels the UIRs of the universal cover of E(2) and
$\eta$ is real. (In this expression $f,g$ are
trigonometric polynomials. However, these quantizations can be
extended to the case when
$f,g$ are arbitrary smooth functions on $S^1$.) Since $P^1$ is
maximal (this is proven below), Corollary~\ref{cor:dne} implies
that none of these quantizations can be extended beyond $P^1$ in
$P.$ 

We now classify the maximal subalgebras of $P$ containing $\bb$.
First, we have
\begin{prop}
$P^1$ is a proper maximal Poisson subalgebra of $P$.
\label{prop:max}
\end{prop}

We need a few preliminaries. Notice that we may equally well view
$P$ as consisting of sums of multiples of terms of types 

\[\ell^{\,r}\sin m\T,\;\;\ell^{\,r}\cos n\T, \,\mbox{ and }\;
\ell^{\,r}\sin m\T \cos n\T.\]

\noi For each integer
$k$ define endomorphisms
$C_k$ and
$S_k$ of each $P_r$ by
\[C_k(p) = \{\ell \cos k\T,p\}\;\;\;\mbox{and}\;\;\;S_k(p) = \{\ell
\sin k\T,p\}.\]

\begin{lem} Each $P_r$ is irreducible under the endomorphisms
$\{C_k,S_k\,|\,k \in \z\}$.
\label{lem:irred}
\end{lem}

\noi {\em Proof:} Let $S$ be an invariant subspace of $P_r$. Then
its complexification $S_{\c} \subset (P_r)_{\c}$ is invariant under
the ladder endomorphisms $L_k = C_k + iS_k,\; k \in \z$. Set $e_m^r
:= \ell^{\, r} e^{im\T}$; then \begin{equation} L_k(e_m^r) = i(m -
kr)e_{m+k}^r. \label{eq:ladder}
\end{equation}

\noi We will show that $S_{\c} = (P_r)_{\c}$, whence $S = P_r$. 

We assert that $S_{\c}$ contains a monomial. Indeed, given $p
\in S_{\c}$ we may write
\[p = \sum_m a_m e_m^r.\]

\noi Let $M$ be any integer such that $a_M \neq 0$. Since $S_{\c}$
is invariant, it follows from (\ref{eq:ladder}) that if $p$ is not
already a monomial, then
$p' = L_0(p) - iMp$ is nonvanishing and belongs to $S_{\c}$. But
$p'$ has one fewer term than $p$. Applying this procedure (which we
refer to as the ``elimination trick'') to
$p'$ and continuing in this fashion, we eventually produce a
monomial $e_N^r \in S_{\c}$.

Then, by successively applying the ladder endomorphisms $L_k$ to
$e_N^r$ for various values of $k$, it is readily verified that we
can obtain any other monomial $e_{N'}^r \in S_{\c}$.
$\bigtriangledown$ 

\ms

\noi {\em Proof of Proposition} \ref{prop:max}: Let $p \not \in
P^1$, and let $R$ be the Poisson algebra generated by $p$ along
with $P^1$. The degree of $p$ in $\ell$ is $r > 1$. By bracketing
$p$ with
$\cos
\T$ a total of $r-2$ times, we obtain an element of $R$ which is
quadratic in $\ell$. Subtracting off the affine terms in
$\ell$--which belong to $P^1$--we obtain an element of ${R} \cap
P_2$. Since both $\ell \sin k\T$ and $\ell \cos k\T$ belong to
$P^1$, Lemma~\ref{lem:irred} implies that $P_2 \subset R.$ Since
$\{\ell^{\, n},\ell^{\, 2} \cos \T\} \in {R} \cap P_{n+1}$, Lemma~
\ref{lem:irred} and induction yield
$P_{n+1}
\subset R$ for all $n > 1$, whence $P \subseteq R$. $\Box$ 

\vskip 12pt

%%%%%%%%%%%%%%%%%%%%%%%%%%

However, $P^1$ is not the only maximal subalgebra of $P$ containing
$\bb$. For each real number $\alpha$, let $W_{\alpha}$ be the
subalgebra of $P_{\c}$ generated by
\[\big\{1, e^0_{\pm 1},e^1_0, e^2_{2N+1} + 2\alpha e^1_{2N+1}\,|\,N
\in \z\big\},\]

\noi where as before $e_m^r := \ell^{\, r} e^{im\T}$. By
construction
$W_{\alpha}$ is totally real (i.e., $\overline {W_{\alpha}} =
W_{\alpha}$), so it must be the complexification of a real
subalgebra $V_{\alpha}.$ That each $V_{\alpha}$ is proper and
maximal is established during the proof of
Proposition~\ref{prop:Vmax} below. 

First, we state a structural result concerning $\wa$, which follows
from a consideration of Poisson brackets of elements of the form
given above. \begin{lem} For each $r$ and $N$ of opposite parity,
there exists an element of $\wa$ of the form
\[ e^{r}_{N} + r\alpha e^{r-1}_{N} + \mbox{\rm l.d.t.,} \]

\noi where\/ {\rm ``l.d.t.''} stands for lower degree terms in
$\ell$. \label{lem:str}
\end{lem}

With this observation, we can now prove
\begin{prop} $P^1$ and $V_{\alpha}, \alpha \in \r$, are the only
proper maximal Poisson subalgebras of
$P$ containing $\bb$.
\label{prop:Vmax}
\end{prop}

\noi {\em Proof:} Since all algebras under consideration here are
real, we can manipulate their complexifications and then take real
parts. It therefore suffices to prove that
$P^1_{\c}$ and
$\wa$ are the only proper maximal totally real Poisson subalgebras
of $P_{\c}$ containing 
\[\bb_{\c} = \sp\big\{1,e^0_{\pm 1},e^1_0 \big\}.\] 

\noi The proof will proceed in
several steps. 

\vskip 12pt

{\em Step 1:} Let $S\subset P$ be a proper maximal subalgebra
containing $\bb$. If $S_{\c} \subset P^1_{\c}$ strictly, then
$S_{\c}$ can't be maximal by Proposition~\ref{prop:max}. Thus
either $S_{\c} = P^1_{\c}$ or $S_{\c} \not \subseteq P^1_{\c}$. 

Suppose $S_{\c} \not \subseteq P^1_{\c}$, and consider any element
of $S_{\c}$ which does not belong to
$P^1_{\c}$. It must have degree at least 2 in $\ell$. By repeatedly
bracketing it with $e_{-1}^0 \in \bb_{\c}$, we obtain an element of
$S_{\c} \cap P^2_{\c}$ of the form
\[\sum_m a_me_m^2 + \mbox{ l.d.t.}\]

\noi with $a_m \neq 0$ for at least one value of $m$. Since $L_0$
preserves $S_{\c}$ we may, by virtue of the elimination trick from
the proof of Lemma~\ref{lem:irred}, suppose that the coefficient of
$\ell^{\, 2}$ in this expression is a trigonometric monomial. Thus
we have
\[ p := e_M^2 + \mbox{ l.d.t.} \in S_{\c} \] 

\noi for some fixed integer $M$. Since $S_{\c}$ is totally real,
$\bar p \in S_{\c}$ as well.

Now compute
\[\{p,e^0_{\pm 1}\} = \pm 2ie_{M \pm 1}^1 + \mbox{ l.d.t.} \in
S_{\c}.\] 

\noi By bracketing this expression with $\bar p$, we find that
$p_{\pm 1} := e^2_{\pm 1} +
\mbox{ l.d.t.}
\in S_{\c}.$ Bracketing $p_{\pm 1}$ with $e^0_{\pm 1}$, we
find that $e^1_{\pm 2} + \mbox{ l.d.t.}$ belongs to $S_{\c}$.
Further bracketing this last expression with $p_{\pm 1}$, we find
that $p_{\pm 3} := e^2_{\pm 3} +
\mbox{ l.d.t.}
\in S_{\c}$. Bracketing $p_{\pm 3}$ with $e^0_{\pm 1}$, we find
that $e^1_{\pm 4} + \mbox{ l.d.t.}$ belongs to $S_{\c}$. Continuing
in this manner, we conclude that
\begin{equation} e^2_{2N+1} + \mbox{ l.d.t.} \in S_{\c}
\label{eq:quads}
\end{equation}

\noi and
\begin{equation} e^1_{2N} + \mbox{ l.d.t.} \in S_{\c} \label{eq:2n}
\end{equation}
\noi for all integers $N.$ Furthermore, by bracketing (\ref{eq:2n})
with $e^0_{\pm 1}$, we have that
\begin{equation} e^0_{2N+1} \in S_{\c}
\label{eq:2n+1}
\end{equation}

\noi for all integers $N.$

\vskip 12pt

{\em Step 2:} We examine (\ref{eq:quads}) and (\ref{eq:2n}) more
closely; we claim that these can be improved as follows: There
exists a real number $\alpha$ such that
\begin{equation} e^2_{2N+1} + 2\alpha e^1_{2N+1}\in S_{\c}
\label{eq:quads=}
\end{equation}

\noi and
\begin{equation} e^1_{2N} + \alpha e^0_{2N} \in S_{\c}
\label{eq:2n=}
\end{equation}
\noi for each integer $N$, respectively. 

First, we note the following useful result, which we refer to as
the ``bootstrap trick.'' Suppose that
$e^0_{2K} \in S_{\c}$ for some
$K
\neq 0$. By bracketing
$e^0_{2K}$ with (\ref{eq:2n}) we see that $e^0_{2K + 2N} \in
S_{\c}$ for all integers
$N$. When combined with (\ref{eq:2n+1}), this implies $P^0 \subset
S_{\c}$, and then (\ref{eq:2n}) yields $e^1_{2N} \in S_{\c}$ for
all $N$, i.e., $(P_1)_{\c} \subset S_{\c}$. But altogether this
implies $P^1_{\c} \subset S_{\c}$, and Proposition~\ref{prop:max}
then forces $S_{\c} = P^1_{\c}$, contrary to assumption. Thus for
no $N \neq 0$ can $e^0_{2N}$ belong to $S_{\c}$.

If every $e^1_{2N} \in S_{\c}$, then (\ref{eq:2n=}) holds with
$\alpha = 0.$ So suppose that
$e^1_{2L}
\not
\in S_{\c}$ for some $L \neq 0.$ Without loss of generality, we may
assume that $L > 0$.\footnote{Since $S_{\c}$ is totally real, if
$e^1_{2L} \not
\in S_{\c}$ then its conjugate $e^1_{-2L} \not
\in S_{\c}$ either.} Then by virtue of (\ref{eq:2n}), and taking
into account (\ref{eq:2n+1}), there must exist a polynomial
\begin{equation} f^1_{2L} = e^1_{2L} + \sum_{K \neq
0}\alpha_{L,K}e^0_{2K} \in S_{\c}, \label{eq:f1}
\end{equation}

\noi where $\alpha_{L,K} \neq 0$ for at least one value of $K$.
Eliminating the top term in $f^1_{2L}$, we obtain: \[f^0_{2L} :=
L_0\big(f^1_{2L}\big) - i2Lf^1_{2L} = 2i \sum_{K \neq 0}
\alpha_{L,K}(K - L)e^0_{2K} \in S_{\c}.\] 

\noi If $\alpha_{L,K} \neq 0$ for some $K \neq L$, we may again use
the elimination trick to remove every term in $f^0_{2L}$ except the
one corresponding to this value of $K$, thereby obtaining $e^0_{2K}
\in S_{\c}$. The bootstrap trick then leads to a contradiction
unless $\alpha_{L,K} = 0$ for all $K \neq L$, in which case
(\ref{eq:f1}) reduces to
\begin{equation} f^1_{2L} = e^1_{2L} + \alpha_Le^0_{2L} \in S_{\Bbb
C}, \label{eq:f11}
\end{equation}

\noi where $\alpha_L := \alpha_{L,L} \neq 0$. We remark that
$\alpha_L$ is uniquely determined by $L$. (Otherwise, upon
subtracting two such
$f^1_{2L}$, we would obtain $e^0_{2L} \in S_{\c}$, which would
again lead to a contradiction.) 

Now consider the quadratic term $e^2_{2L+1}$. It cannot belong to
$S_{\c}$, since otherwise
$\big\{e^2_{2L+1},e^0_{-1}\big\} = -2ie^1_{2L}$ would also. {}From
(\ref{eq:quads}), then,
\[f^2_{2L+1} = e^2_{2L+1} + \sum_K \beta_{L,K} e^1_{2K+1} + \mbox{
l.d.t.} \in S_{\c},\]

\noi where we made use of (\ref{eq:2n}). Now
\[\big\{f^2_{2L+1},e^0_{-1}\big\} = -2i\Big(e^1_{2L} + {\textstyle
\frac{1}{2}}\sum_K \beta_{L,K}e^0_{2K} \Big) \in S_{\c}\] 

\noi and comparison with (\ref{eq:f11}) along with the bootstrap
trick gives
$\beta_{L,K} = 2\alpha_L\delta_{K,L}.$ Thus \[f^2_{2L+1} =
e^2_{2L+1} + 2\alpha_L e^1_{2L+1} + \mbox{ l.d.t.}\]

\noi But then
\[\big\{f^2_{2L+1},e^0_1\big\} = 2i \big(e^1_{2L+2} + \alpha_L
e^0_{2L+2}\big) \in S_{\c}.\]

\noi If $e^1_{2L+2} \in S_{\c}$ then, to avoid a contradiction via
the bootstrap trick, we must have $\alpha_L = 0,$ which is
impossible. Thus
$e^1_{2L+2}
\not
\in S_{\c}$, in which case comparison with (\ref{eq:f11}) yields
$\alpha_{L + 1} =
\alpha_L$.

A similar argument using $e^2_{2L-1}$ in place of $e^2_{2L+1}$
yields $e^1_{2L-2}
\not
\in S_{\c}$ along with
$\alpha_{L - 1} =
\alpha_L$, provided $L \neq 1$. Iterating, we obtain $e^1_{2N} +
\alpha_L e^0_{2N} \in S_{\c}$ for all positive integers $N$.
Analogously, starting with the conjugate $e^1_{-2L}$ of $e^1_{2L}$
(recall that $S_{\c}$ is totally real), we obtain $e^1_{2N} +
\alpha_{-L} e^0_{2N} \in S_{\c}$ for all negative integers $N$.
Comparing the bracket \[\big\{f^2_{2L+1},e^0_{-4L - 1}\big\} =
-2i(4L+1)\big(e^1_{-2L} + \alpha_{L}e^0_{-2L}\big) \in S_{\c}\]

\noi with (\ref{eq:f11}) and applying the bootstrap trick, we get
$\alpha_{-L} = \alpha_L.$ Thus (\ref{eq:2n=}) holds for all
integers $N$ with $\alpha := \alpha_L.$ Finally, comparing the
conjugate of $f^1_{2L}$ with $f^1_{-2L}$ gives
${\overline {\alpha_L}} = \alpha_{-L} = \alpha_L$, so $\alpha$ is
real.

It is now a simple matter to prove (\ref{eq:quads=}). {}From the
arguments above coupled with (\ref{eq:2n+1}), we may write
\[f^2_{2N+1} = e^2_{2N+1} + 2\alpha e^1_{2N+1} + \sum_K
\tau_{N,K}e^0_{2K} \in S_{\c}\] 

\noi for all integers $N$ and some coefficients $\tau_{N,K}.$ But
now the elimination trick gives
\[i\sum_K (2K - 2N -1)\tau_{N,K}e^0_{2K} \in S_{\c},\] 

\noi which leads to a contradiction unless $\tau_{N,K} = 0$ for all
$K$.

%%%%%%%%%%%%%%%

\vskip 12 pt

{\em Step 3:} We first observe that according to (\ref{eq:quads=})
\begin{equation}
\wa \subseteq S_{\c}
\label{eq:cap}
\end{equation}

\noi for some $\alpha \in \r$. We will show that $S_{\c } \subseteq
\wa$. 

Let $q \in S_{\c}$ be of degree $r$ in $\ell$, and suppose that $q
\not \in \wa.$ By Lemma~\ref{lem:str}, we know that $e^r_N +
r\alpha e^{r-1}_{N} + \cdots \in
\wa$ for all $N$ with opposite parity to $r$, and so by
(\ref{eq:cap}) we may eliminate all such combinations in $q$,
thereby obtaining a polynomial $\tilde q$ which belongs to $S_{\c}$
but not $\wa$. Now either we can eliminate all terms of degree
$r$ in this manner, in which case $\tilde q$ has degree $\tilde r
\leq r-1$, or else there is a term in
$\tilde q$ of the form
$e^r_N$ where
$N$ and
$r$ have the same parity. In the latter instance, we may isolate
this term using the elimination trick, and then bracket with
$e^0_{-1}$ $r$ times to obtain $e^0_{N-r} \in S_{\c}.$ Since $N$
and $r$ have the same parity, $N-r$ is even. If $N-r \neq 0$, then
the bootstrap trick produces a contradiction. If $N = r$, then
bracket instead with $e^0_1$ to obtain $e^0_{N+r} \in S_{\c}$ with
$N+r$ even and nonzero. 

We iterate this procedure, either encountering a contradiction at
some point or finally ending up with a zeroth degree polynomial
$\tilde q' \in S_{\c}$ of the form
$\sum_N \gamma_N e^0_{2N}$. Since $\tilde q' \not \in \wa$,
$\gamma_M
\neq 0$ for some
$M \neq 0$. But the elimination trick can now be used to produce
$e^0_{2M}
\in S_{\c}$, which also yields a contradiction. Thus in all
eventualities, the assumption that $q \not \in \wa$ produces a
contradiction. It follows that
$S_{\c} = W_{\alpha}$. $\Box$

\ms In fact, other than $P^1$ and subalgebras thereof, this proof
shows that the $\va$ are the {\em only\/} proper subalgebras of $P$
containing $\bb$.

\ms

In contrast to
$P^1$, we now show that there is {\em no\/} nontrivial quantization
of $\va$ which represents $\bb$ irreducibly. While the method of
proof is the same as that of the no-go theorem for $P$ in \S2, we
must make sure that all constructions take place in $\va$. Upon
replacing the identity (\ref{eq:iden}) by
\begin{equation}
\big\{\{\ell^{\, 3} + 3\alpha \ell^{\,2},\sin \T\},\sin \T\big\} +
\big\{\{\ell ^{\, 3} + 3\alpha \ell^{\,2},\cos \T\},\cos \T\big\} =
6\ell + 6\alpha,
\label{eq:newiden}
\end{equation}

\noi the proof of Proposition~\ref{prop:l2} can be immediately
adapted to give
\begin{prop}
$\q(\ell^{\,3} + 3\alpha \ell^{\,2}) = \q(\ell)^3 + 3\alpha
\q(\ell)^2 + b'\q(\ell) + c'I$, where
$b',c' \in \r$ are arbitrary.
\label{prop:cubicvn}
\end{prop}

This can be specialized further: Quantizing the bracket relations \[
(\ell^{\,2} + 2\alpha \ell) \sin \T = -\frac{1}{3}\{\ell^{\, 3} +
3\alpha \ell^{\,2},\cos \T\} \nonumber \label{eq:pb1}
\]

\noi and
\[ (\ell^{\,2} + 2\alpha \ell) \cos \T =
\frac{1}{3}\{\ell^{\, 3} + 3\alpha
\ell^{\,2},\sin \T\} \nonumber
\label{eq:pb2}
\]

\noi we obtain\footnote{These calculations were done using the {\sl
Mathematica} package {\sl NCAlgebra} \cite{HM}.} \begin{eqnarray}
\q\big((\ell^{\,2} + 2\alpha \ell) \sin \T\big) & = & \q(\sin
\T)\q(\ell)^2 + \big(2\alpha \q(\sin \T) - i\q(\cos
\T)\big)\q(\ell) \nonumber
\\ & & \mbox{} + \frac{1+b'}{3}\q(\sin
\T) - i\alpha \q(\cos \T)
\label{eq:l2s'}
\end{eqnarray}

\noi and
\begin{eqnarray}
\q\big((\ell^{\,2} + 2\alpha \ell) \cos \T\big) & = & \q(\cos
\T)\q(\ell)^2 + \big(2\alpha \q(\cos \T) + i\q(\sin
\T)\big)\q(\ell) \nonumber \\ & & \mbox{} + \frac{1+b'}{3}\q(\cos
\T) + i\alpha \q(\sin \T). \label{eq:l2c'}
\end{eqnarray}

\noi Using these to quantize
\[\ell^{\,3} + 3\alpha \ell^{\,2} =
\frac{1}{2}\big\{(\ell^{\,2} + 2\alpha \ell) \cos \T,(\ell^{\,2} +
2\alpha \ell)\sin \T\big\} - 2\alpha^2 \ell,\] 

\noi we get
\[\q(\ell^{\,3} + 3\alpha \ell^{\,2}) = \q(\ell)^3 + 3\alpha
\q(\ell)^2 + \frac{1+b'}{3}\q(\ell) + \frac{1+b'}{3}\alpha \, I,\] 

\noi which is compatible with Proposition~\ref{prop:cubicvn} iff
$b' = \frac{1}{2}$ and $c' =
\frac{\alpha}{2}$. Thus,
\[\q(\ell^{\,3} + 3\alpha \ell^{\,2}) = \q(\ell)^3 + 3\alpha
\q(\ell)^2 + \frac{1}{2}\q(\ell) + \frac{\alpha}{2}I.\] 

\noi (As an aside, observe that fixing $b' = \frac{1}{2}$ here
leads to an inconsistency with our calculations in \S2, where $b =
0$. Indeed, comparing the expressions (\ref{eq:l2s})$\mbox{ }\! +
\,2\alpha \times \!\!\mbox{ }$(\ref{eq:ls}) with (\ref{eq:l2s'})
for $\q\big((\ell^{\,2} + 2\alpha \ell) \sin \T\big)$, and
(\ref{eq:l2c})$\mbox{ }\! + \,2\alpha \times \!\!\mbox{
}$(\ref{eq:lc}) with (\ref{eq:l2c'}) for
$\q\big((\ell^{\,2} + 2\alpha \ell)
\cos \T\big)$, we see that they differ in the zeroth degree terms
in $\q(\ell)$. We could equally well have used this discrepancy as
a basis for the previous no-go result.)

Finally, using these \vn\ rules, we quantize \[(\ell^{\, 4} +
4\alpha \ell^{\,3} + 4 \alpha^2 \ell^{\,2}) \sin \T
=\frac{1}{3}\big\{(\ell^{\,2} + 2\alpha \ell) \cos
\T,\ell^{\, 3} + 3 \alpha \ell^{\,2}\big\}\] \noi and
\[(\ell^{\, 4} + 4\alpha \ell^{\,3} + 4 \alpha^2 \ell^{\,2}) \cos
\T = -\frac{1}{3}\big\{(\ell^{\,2} + 2\alpha \ell) \sin \T,\ell^{\,
3} + 3 \alpha \ell^{\,2}\big\}\] 

\noi to get
\begin{eqnarray}
\q\big((\ell^{\, 4} + 4\alpha \ell^{\,3}\hspace{-2ex} & + &
\hspace{-2ex}4\alpha^2
\ell^{\,2}) \sin \T\big) \nonumber \\ & \mbox{} = & \q(\sin
\T)\q(\ell)^4 \nonumber \\ & & \mbox{} + \big(4\alpha \q(\sin \T)
-2i\q(\cos \T)\big)\q(\ell)^3 \nonumber \\ & &\mbox{} +
\big([4\alpha^2 + 2] \q(\sin \T) - 6i\alpha \q(\cos
\T)\big)\q(\ell)^2 \nonumber \\ & & \mbox{} +
\big(4\alpha \q(\sin \T) -i[4 \alpha^2 + 1]\q(\cos \T)\big)\q(\ell)
\nonumber
\\ & & \mbox{} +
\big({\textstyle \frac{1}{4}} + \alpha^2\big)\q(\sin \T) -i\alpha
\q(\cos
\T)
\label{eq:l4s}
\end{eqnarray}

\noi and
\begin{eqnarray}
\q\big((\ell^{\, 4} + 4\alpha \ell^{\,3} \hspace{-2ex} & + &
\hspace{-2ex} 4\alpha^2
\ell^{\,2}) \cos \T\big) \nonumber \\ & \mbox{} = & \q(\cos
\T)\q(\ell)^4 \nonumber \\ & & \mbox{} + \big(4\alpha \q(\cos \T) +
2i\q(\sin \T)\big)\q(\ell)^3 \nonumber \\ & & \mbox{} +
\big([4\alpha^2 + 2] \q(\cos \T) + 6i\alpha \q(\sin
\T)\big)\q(\ell)^2 \nonumber \\ & & \mbox{} +
\big(4\alpha \q(\cos \T) + i[4 \alpha^2 + 1]\q(\sin
\T)\big)\q(\ell) \nonumber \\ & & \mbox{} + \big({\textstyle
\frac{1}{4}} + \alpha^2\big)\q(\cos \T) + i\alpha
\q(\sin
\T).
\label{eq:l4c}
\end{eqnarray}

We are now ready for:

\begin{thm} There is no nontrivial quantization of $(\va,\bb)$.
\label{thm:Vnogo}\end{thm}

\noi {\em Proof:} We consider representations of $\bb$ of
type ({\em i\/})$'$, and use the
\vn\ rules (\ref{eq:l2s'})--(\ref{eq:l4c}) to quantize the bracket
relation
\begin{eqnarray*}
\lefteqn{\hspace{-4ex}\left\{\big\{(\ell^{\, 2} + 2\alpha \ell)
\cos \T,(\ell^{\, 4} + 4\alpha
\ell^{\,3} + 4\alpha^2 \ell^{\,2})\sin \T\big\}, \cos
\T\right\} } \\ & \mbox{} + &\left\{\big\{(\ell^{\, 4} + 4\alpha
\ell^{\,3} + 4\alpha^2 \ell^{\,2}) \cos \T,(\ell^{\, 2} + 2\alpha
\ell) \sin \T\big\},
\cos \T\right\} \\ & & \mbox{} = -30(\ell^{\, 4} + 4\alpha
\ell^{\,3} + 4\alpha^2 \ell^{\,2}) \sin \T - 24\alpha^2(\ell^{\,2}
+ 2\alpha \ell)\sin \T.
\end{eqnarray*}

\noi After another computer calculation, the left hand side reduces
to
\begin{eqnarray*}
\lefteqn{\hspace{-3ex} - 30\q(\sin \T)\q(\ell)^4 + \big(60i\q(\cos
\T) - 120\alpha
\q(\sin \T)\big)\q(\ell)^3\ } \\ & \hspace{10ex} + & \big(180i
\alpha \q(\cos \T) -[84 + 144\alpha^2]\q(\sin \T)\big)\q(\ell)^2 \\
& \hspace{10ex} + & \big(i[54 + 144\alpha^2]\q(\cos \T) -
[168\alpha + 48 \alpha^3]\q(\sin \T)\big)\q(\ell) \\ &
\hspace{10ex} + & i[54 \alpha + 24 \alpha^3]\q(\cos \T)
-\big[{\textstyle \frac{31}{2}} + 66 \alpha^2\big]\q(\sin \T),
\end{eqnarray*}

\noi which is quite different than $-30 \;\times$
(\ref{eq:l4s})$\mbox{ } \! - \,24\alpha^2 \;\times$
(\ref{eq:l2s'}): \begin{eqnarray*}
\lefteqn{\hspace{-3ex} - 30\q(\sin \T)\q(\ell)^4 + \big(60i\q(\cos
\T) - 120 \alpha
\q(\sin \T)\big)\q(\ell)^3\ } \\ & \hspace{10ex} + & \big(180i
\alpha \q(\cos \T) -[60 + 144\alpha^2]\q(\sin \T)\big)\q(\ell)^2 \\
& \hspace{10ex} + & \big(i[30 + 144\alpha^2]\q(\cos \T) -
[120\alpha + 48 \alpha^3]\q(\sin \T)\big)\q(\ell) \\ &
\hspace{10ex} + & i[30 \alpha + 24 \alpha^3]\q(\cos \T)
-\big[{\textstyle \frac{15}{2}} + 42 \alpha^2\big]\q(\sin
\T).\;\;\Box
\end{eqnarray*}

On the other hand, there do exist trivial, but nonetheless nonzero,
representations of type ({\em ii\/})$'$, provided $\alpha \neq 0$.
To see this, quantize (\ref{eq:newiden}) to obtain $\q(\ell) =
-\alpha I$. Moreover, since $\big\{e^{3}_{2N} + 3\alpha e^2_{2N} -
2\alpha^3 e^0_{2N},e^0_1\big\} = 3i\big(e^{2}_{2N+1} + 2\alpha
e^1_{2N+1}\big)$, quantization yields $\q\big(e^{2}_{2N+1} +
2\alpha e^1_{2N+1}\big) = 0$ for all $N$. It follows from the
definition of $\wa$ that the only observables in $\va$ which have
nonzero quantizations are of of the form $b\ell + c$, with
$\q(b\ell + c) = (c - \alpha b)I.$ This is reminiscent of the
situation for
$S^2$, cf. \cite{GGH}.

Thus the largest quantizable subalgebra of $P$ containing $\bb$ is
$P^1$. At the beginning of this section, we exhibited certain
quantizations $\q_{\nu,\eta}$ of $(P^1,\bb)$. In fact, as we now
show, these are the only ones.
\begin{thm} If $\q$ is a nontrivial quantization of $(P^1,\bb)$,
then $\q = \q_{\nu,\eta}$ for some
$\nu \in [0,1)$ and $\eta \in \r$.
\label{thm:unique}
\end{thm}

\noi {\em Proof:} We may suppose that $\q$ restricted to $\bb$ is
given by ({\em i\/})$'$ for some $\nu$. In what follows it is
convenient to use complex notation.

Because of the linearity of $\q$, to establish (\ref{eq:qnu}) it
suffices to prove that
\begin{equation}
\q(e^{iN\T}) = e^{iN\T}
\label{eq:qf}
\end{equation}

\noi and
\begin{equation}
\q\big(\ell \, e^{iN\T}\big) = e^{iN\T}\left(-i \frac{d}{d\T} +
\left[iN\eta + \frac{N}{2} + \nu\right]\!I\right). \label{eq:qN}
\end{equation} 

We begin by quantizing the bracket relation
$\big\{\ell,e^{iN\T}\big\} = iNe^{iN\T}$ to get
$\big[\q(\ell),\q(e^{iN\T})\big] = N\q(e^{iN\T})$. Writing
$\q(e^{iN\T})|n\rangle = \sum_k D^N_{nk}|k\rangle$, we evaluate the
matrix element \[\left \langle m \big |
\big[\q(\ell),\q(e^{iN\T})\big] \big | n\right \rangle = N \left
\langle m \big | \q(e^{iN\T}) \big | n\right \rangle\]

\noi to obtain
$mD^N_{nm} - nD^N_{nm} = ND^N_{nm}$, i.e., \[(m-n-N)D^N_{nm} =
0.\]

\noi Thus $D^N_{nm} = 0$ unless $m=n+N$, and hence
\begin{equation}
\q(e^{iN\T})|n\rangle = D^N_n|n+N\rangle, \label{eq:D}
\end{equation}

\noi where we have abbreviated $D^N_{n,n+N} =: D^N_n.$ Note in
particular that $D^0_n = 1$ and $D^{\pm 1}_n = 1,$ for all $n$, cf.
({\em i\/})$'$. A similar analysis of the relation
$\big\{\ell,\ell \, e^{iN\T}\big\} = iN\ell \,e^{iN\T}$ yields
\begin{equation}
\q(\ell \, e^{iN\T})|n\rangle = d^N_n|n+N\rangle, \label{eq:d}
\end{equation}

\noi where in particular $d^{\,0}_n = n + \nu$ by virtue of ({\em
i\/})$'$.

Next quantize $\big\{e^{iN\T},e^{i\T}\big\} = 0$ using ({\em
i\/})$'$ to obtain
$\big[\q(e^{iN\T}),e^{i\T}\big] = 0$. Applying this to a ket
$|n\rangle$, (\ref{eq:D}) gives
\[(D^N_{n+1} - D^N_{n})|n + N + 1\rangle = 0\] 

\noi for all integers $n$, from which we conclude that $D^N_n$
depends only upon
$N$, and so will be denoted $D^N$ henceforth. Thus \begin{equation}
\q(e^{iN\T}) = D^Ne^{iN\T}.
\label{eq:qD}
\end{equation}

Now we quantize
\[\big\{\ell \, e^{iN\T},e^{iM\T}\big\} = iMe^{i(N+M)\T}\] 

\noi to obtain
\[
\big[\q(\ell \, e^{iN\T}),\q(e^{iM\T})\big] =
M\q\big(e^{i(N+M)\T}\big). \]

\noi Applying this to a ket $|n\rangle$, (\ref{eq:qD}) and
(\ref{eq:d}) give
\[\big(d^N_{n+M} - d^N_n\big)D^M|n+M+N \rangle = MD^{N+M}|n+M+N
\rangle,\] 

\noi from which we conclude that
\begin{equation}
\big(d^N_{n+M} - d^N_n\big)D^M = MD^{N+M}. \label{eq:MN}
\end{equation}

Setting $M = 1$ this reduces to $d^N_{n+1} - d^N_{n} = D^{N+1}$,
which in turn implies that
\begin{equation} d^N_{n} = d^N_0 + nD^{N+1}.
\label{eq:dnD}
\end{equation}

\noi On the other hand, taking $M = -N$, (\ref{eq:MN}) reduces to
\[\big(d^N_{n-N} - d^N_n\big)D^{-N} = -N.\] 

\noi Substituting (\ref{eq:dnD}) into this, we find that
$D^{N+1}D^{-N} = 1$ for all integers $N$. Since $D^{\pm 1} = 1$,
this implies that each $D^N = 1$. Thus (\ref{eq:qf}) is proved. 

Following the established pattern, upon quantizing the relation
\[\big\{\ell \, e^{iN\T},\ell \, e^{iM\T}\big\} =
i(M-N)\,\ell\,e^{i(M+N)\T}\]

\noi and using (\ref{eq:dnD}), we eventually produce
\begin{equation} Md^M_0 - Nd^N_0 = (M-N)d^{M+N}_0.
\label{eq:recur1}
\end{equation}

\noi Taking $M = -N$, this becomes
\begin{equation} d^N_0 + d_0^{-N} = 2\nu.
\label{eq:plus}
\end{equation}

\noi Taking $M = \pm 1$, (\ref{eq:recur1}) reduces to \[d^1_0 -
Nd_0^N = (1-N)d_0^{N+1} \]

\noi and
\[d^{-1}_0 + Nd_0^N = (1+N)d_0^{N-1}.\]

\noi Relabeling $N \mapsto N + 1$ in this last equation, and
then eliminating $d_0^{N+1}$ between these two equations using
(\ref{eq:plus}) gives the relation
\begin{equation} d_0^N = Nd^1_0 +(1-N)\nu.
\label{eq:recur2}
\end{equation}

Furthermore, we observe that by the definition of a quantization
\[\q\big(\ell \, e^{-iN\T}\big) \subset \q\big(\ell \,
e^{iN\T}\big)^*,\] 

\noi which in particular forces
\begin{equation}
\overline{d^N_0} - d^{-N}_0 = N.
\label{eq:minus}
\end{equation}

\noi Adding (\ref{eq:plus}) and (\ref{eq:minus}), we get $\Re
\big(d_0^N \big) = \frac{N}{2} + \nu$. {}From (\ref{eq:recur2})
and its conjugate, we obtain $\Im \big(d^N_0\big) = N \Im
\big(d_0^1\big) =: N\eta.$ Substituting back into (\ref{eq:d}) and
recalling (\ref{eq:dnD}), we end up with 
\[ \q\big(\ell \, e^{iN\T}\big)|n\rangle = \left(n + iN\eta + 
{\textstyle \frac{N}{2}} + \nu \right)|n+N\rangle,\] 

\noi which is equivalent to (\ref{eq:qN}). $\Box$ 

\ms

Thus within the subalgebra of polynomials, the quantizations
$\q_{\nu,\eta}$ of
$(P^1,\bb)$ are the best one can do.

\end{section}

%%%%%%%%%%%%%%%%%%%%%%%%%%%%%%%%%%%%%%%%%%%%%%%%%%%%%%%%%%%%%%%%%%%%%%
%%%%%%%%%%%%%%%%%%%%%%%%%%%%%%%%%%%%%%%%%%%%%%%%%%%%%%%%%%%%%%%%%%%%%%
%%%%%%%%%%%%%%%%%%%%%%%%%%%%%%%%%%%%%%%%%%%%%%%%%%%%%%%%%%%%%%%%%%%%%%
%%%%%%%%%%%%%%%%%%%%%%%%%%%%%%%%%%%%%%%%%%%%%%%%%%%%%%%%%%%%%%%%%%%%%%
%%%%%%%%%%%%%%%%%%%%%%%%%%%%%%%%%%%%%%%%%%%%%%%%%%%%%%%%%%%%%%%%%%%%%%
%%%%%%%%%%%%%%%%%%%%%%%%%%%%%%%%%%%%%%%%%%%%%%%%%%%%%%%%%%%%%%%%%%%%%% 

\begin{section}{Discussion}

Although for topological and algebraic reasons the quantization of
$T^*\!S^1$ might be expected to share some of the features of both
those of $\r^2$ and
$T^2$, we see that in all essential respects it behaves like the
plane. For both
$\r^2$ and $T^*\!S^1$ there is an obstruction, and a maximal
subalgebra of polynomial observables that can be consistently
quantized consists of those polynomials which are affine in the
momentum. Most likely, the underlying reason is that in these
examples the given basic sets are the generators of transitive
(finite-dimensional) Lie group actions \big(the Heisenberg group
H(2) for $\r^2$, and the Euclidean group E(2) for $T^*\!S^1$\big),
whereas this is not true for the basic sets $\bb_k$ on the torus.

There are some differences, however, which reflect the non-simple
connectivity of $T^*\!S^1$. For instance, on $\r^2$, there are
exactly three maximal polynomial subalgebras containing the basic
set
$\sp\{1,q,p\}$, whereas according to Proposition~\ref{prop:Vmax}
there is an infinity of such containing $\bb$ for the cylinder.
Moreover, on $\r^2$ all three of these maximal subalgebras can be
consistently quantized \cite{GGT}. But on
$T^*\!S^1$, in view of Theorem~\ref{thm:Vnogo} and the discussion
at the beginning of \S3, only one of these can be nontrivially
quantized (viz. $P^1$), leading to the position representations
(\ref{eq:qnu}) on $L^2(S^1)$. Thus there is no analogue of the
metaplectic representation for $T^*\!S^1$. Since
$\T$ is an angular variable, there is also no cylindrical
counterpart of the momentum representation. Another key difference
will be discussed below. Thus the possible polynomial quantizations
of $T^*\!S^1$ are more limited than those of $\r^2$; this is
surprising, given that $T^2$ admits a full quantization.

Although an obstruction exists for the cylinder, as predicted by
the conjecture in \S1, this example serves to disprove part of
another conjecture in \cite{GGT} concerning the maximal subalgebras
of observables that can be consistently quantized. In the present
context, this conjecture can be stated:
\begin{quote}
\em Let\/ $\bb$ be a basic set, which is itself a Poisson
subalgebra of\/ $\p(M)$. Then every integrable irreducible
representation of\/
$\bb$ can be extended to a quantization of
$\big(\nn(\bb),\bb\big)$, where\/ $\nn(\bb)$ is the normalizer of\/
$\bb$ in\/ $\p(M)$. Furthermore, no nontrivial quantization of\/
$\big(\nn(\bb),\bb\big)$ can be extended beyond\/ $\nn(\bb)$.
\end{quote}

\noi For the cylinder, $\nn(\bb) = \bb$. But from \S3, we see that
the representation ({\em i}\/)$'$ can be extended to a quantization
of $({\cal O},\bb)$, where ${\cal O}$ is the subalgebra of
observables which are affine in the (angular) momentum $\ell$. On
$\r^2$, the basic set $\sp\{1,q,p\}$ is not self-normal, and it is
this difference which is largely responsible for the absence of a
``metaplectic-type'' representation on $T^*\!S^1.$

On the other hand, the existence of consistent quantizations of
$({\cal O},\bb)$ can be understood from the standpoint of geometric
quantization theory; since ${\cal O}$ is the normalizer of the
vertical polarization
$\{f(\T)\}$ on $T^*\!S^1$ \cite{Wo}. In fact, the parameter $\nu
\in [0,1)$ in the quantizations $\q_{\nu,\eta}$ labels the
inequivalent connections on the prequantization line bundle $L =
T^*\!S^1 \times \c$. (On the other hand, we do not know if the
parameter $\eta$ in (\ref{eq:qnu})--which to our knowledge appears
here for the first time--has any geometric significance.) While
$\cal O$ thus finds a natural interpretation in the context of
polarizations, it is not at all clear how (or even if) these
quantizations could be predicted by considerations involving $\bb$
alone. An important open problem is therefore to repair this
conjecture.

It is interesting to observe that if we regard $\T$ as a {\em
real\/} variable, then $\bb = \sp\{1, \sin \T,\cos \T, \ell\}$
forms a basic set on $\r^2$ (with coordinates $\T,\ell$); indeed,
$\r^2$ is a Hamiltonian homogeneous space for the universal cover of
E(2).\footnote{Specifically, the action of $\r
\ltimes \r^2$ on $\r^2$ is
\[(t,x,y)\cdot (\T,\ell) = (\T + t, \ell + x\sin(\T + t) - y\cos(\T
+ t)).\]} Thus, if we wish, we may regard $\bb$ as an exotic basic
set on the plane. (The adjective ``exotic'' is perhaps
misleading, as this $\mbox{e(2)} \times \r$ subalgebra on $\r^2$
plays an important role in geometric optics, cf.
\cite[\S17]{GS}.) A moment's reflection shows that the results of
\S2 carry through to this context (with one minor exception, noted
below). Thus we obtain an exotic no-go result for $\r^2$:
\begin{thm} There exists no quantization of
$\big(\p(\r^2),\bb\big).$ \label{thm:r2}
\end{thm}

Comparatively, this result is stronger than \gr's original
no-go theorem \cite{Gr}. Indeed, the latter only states that there
does not exist a quantization of the (standard) polynomial
subalgebra of $\r^2$ which extends the metaplectic representation.
This cannot be strengthened to a statement analogous to
Theorem~\ref{thm:r2} without introducing ad hoc assumptions \`a la
\cite{vH1}. (A fuller discussion of this point can be found in
\cite[\S4.1]{GGT}.) So in fact Theorem~\ref{thm:r2} is the optimal
no-go result extant for $\r^2.$

We make three remarks. First, in
$\p(\r^2)$, $\bb = \sp\{1,\sin \T, \cos \T, \ell\}$ is {\em not\/}
self-normal; in particular, $\T \in \nn(\bb)$. It would be
interesting to discover the ramifications of this, especially as
regards the conjecture above. Second, this basic
set separates points only locally on $\r^2$, not globally,
unlike the Heisenberg basic set on
$\r^2$ or even $\bb$ on $T^*\!S^1$. Third, as described in \S3,
there do exist quantizations of
$(P^1,\bb)$ on $L^2(S^1)$. Curiously, they do not seem to arise
from geometric quantization theory (i.e., via a choice of
polarization on $\r^2$, since $S^1$ cannot be the leaf space of any
foliation of the plane).

Finally, one topic for future exploration would be to consider the
higher-dimensional analogues of the cylinder, viz. $T^*\!S^n$ with
group E($n$).

\end{section}

%%%%%%%%%%%%%%%%%%%%%%%%%%%%%%%%%%%%%%%%%%%%%%%%%%%%%%%%%%%%%%%%%%%%%%
%%%%%%%%%%%%%%%%%%%%%%%%%%%%%%%%%%%%%%%%%%%%%%%%%%%%%%%%%%%%%%%%%%%%%%
%%%%%%%%%%%%%%%%%%%%%%%%%%%%%%%%%%%%%%%%%%%%%%%%%%%%%%%%%%%%%%%%%%%%%%
%%%%%%%%%%%%%%%%%%%%%%%%%%%%%%%%%%%%%%%%%%%%%%%%%%%%%%%%%%%%%%%%%%%%%%
%%%%%%%%%%%%%%%%%%%%%%%%%%%%%%%%%%%%%%%%%%%%%%%%%%%%%%%%%%%%%%%%%%%%%%
%%%%%%%%%%%%%%%%%%%%%%%%%%%%%%%%%%%%%%%%%%%%%%%%%%%%%%%%%%%%%%%%%%%%%% 

%%%%%%%%%%%%%%%%%%%%%%%%%%%%%%%%%%%%%%%%%%%%%%%%%%%%%%%%%%%%%%%%%%%%%%
%%%%%%%%%%%%%%%%%%%%%%%%%%%%%%%%%%%%%%%%%%%%%%%%%%%%%%%%%%%%%%%%%%%%%% 

\end{document}